# Enabling the Analysis of Personality Aspects in Recommender Systems

*Completed Research Paper*

**Shahpar Yakhchi**
Department of Computing
Macquarie University
Sydney, Australia
Shahpar.Yakhchi@hdr.mq.edu.au

**Amin Beheshti**
Department of Computing
Macquarie University
Sydney, Australia
Amin.Beheshti@mq.edu.au

**Seyed Mohssen Ghafari**
Department of Computing
Macquarie University
Sydney, Australia
Seyed-Mohssen.Ghafari@hdr.mq.edu.au
Seyedmohssen.ghafari@data61.csiro.au

**Mehmet Orgun**
Department of Computing
Macquarie University
Sydney, Australia
Mehmet.Orgun@mq.edu.au

## Abstract

*Existing Recommender Systems mainly focus on exploiting users' feedback, e.g., ratings, and reviews on common items to detect similar users. Thus, they might fail when there are no common items of interest among users. We call this problem the Data Sparsity With no Feedback on Common Items (DSW-n-FCI). Personality-based recommender systems have shown a great success to identify similar users based on their personality types. However, there are only a few personality-based recommender systems in the literature which either discover personality explicitly through filling a questionnaire that is a tedious task, or neglect the impact of users' personal interests and level of knowledge, as a key factor to increase recommendations' acceptance. Differently, we identifying users' personality type implicitly with no burden on users and incorporate it along with users' personal interests and their level of knowledge. Experimental results on a real-world dataset demonstrate the effectiveness of our model, especially in DSW-n-FCI situations.*

**Keywords:** Recommender Systems, Personality, User Behavior, User Knowledge.

## Introduction

Recommender Systems (RSs) are decision-making tools and techniques that provide suggestions for items that a particular user is interested in (Ricci, F et.al, 2015). Moreover, in the age of Big Data, many companies have started to trace the activities of customers and capture every click of the mouse in searching, browsing, comparing, as well as the purchasing process. Given this huge amount of data and meta-data generated every second, RSs have an opportunity to make an effective use of big captured data and provide a useful alternative to search algorithms. For example, Collaborative Filtering (CF), i.e., a method of making automatic predictions about the interests of a user by collecting preferences from many users, can highly benefit from this big captured data. CF techniques try to predict the interests of a particular user on items by discovering previously rated similar items, and thus they might fail when there is no captured previous feedback on common items. In this paper, we called this situation the Data Sparsity With no Feedback on Common Items (DSW-n-FCI) problem, which is a subclass of the data sparsity problem when there is not much available information about user-item interactions. Furthermore, in the DSW-n-FCI situation, the existing studies, e.g., CTR (Wang et al. 2011), SVD++ (Koren et al. 2008) might fail, as they are not able to find any connection among users. In such approaches, users' feedback, e.g., ratings and reviews can be the main indicator of identifying similar users and if there is no feedback captured on a common item.

Contrary to such approaches, personality-based RSs do not need to look up for common items of interest to find the similar users, and they can find similar users, even in the presence of DSW-n-FCI, by finding users with similar personality types. Psychologically speaking, personality is a consistent behavior pattern that a person tends to show regardless of her situation. It has also strong correlation with individuals' interests and people with similar personality types tend to share similar interests. This observation has inspired some RSs to employ personality detection methods to boost their performance by identifying users with a similar personality type. However, the main drawback of most of the existing studies is that they highly rely on explicit personality detection approaches such as filling a questionnaire. In these approaches, users may be asked to answer some questions and do a survey which is a time-consuming task and they might be unwilling to participate due to the privacy concerns (Tkalcic et al. 2010, Hu et al. 2010, Nunes 2009, Sitaraman 2015). In addition, people usually do not disclose much about themselves and probably answer questions incorrectly which can have a negative effect on the accuracy of recommender system. Furthermore, there are only a few attempts in the literature to design RSs that detect users' personality types implicitly by analyzing their activities, provided contents, and behaviors. However, they mostly ignore to consider users' personal interests, and their level of knowledge in a particular domain which can affect recommendations' acceptance rate.

In contrast, in one hand, in order to preserve users' actual interests on items and show their personal interests, in this paper we build a user-item interaction matrix with the real value of ratings. On the other hand, we detect users' personality type implicitly with no burden on users through analyzing user-generated contents. We employ the Five Factor Model (FFM) which is one of the widely adopted personality models both in psychology and computer science domains. FFM consists of five main personality traits as Openness to Experience, Conscientiousness, Extraversion, and Neuroticism and it has briefly known as OCEAN. Additionally, it has been discovered in psychology that people with similar personality types, more or less, tend to share similar interests. For instance, people with the Openness personality type mostly prefer to watch comedy and fantasy genre of movies, while romantic movies are more likely to be seen by people with the Neurotic personality type (Cantador et al. 2013). We also adopt one of the leading text analysis tools with a lexical root known as Linguistic Inquiry and Word Count (LIWC) with the ability to discover the words related to more than 88 psychological categories relevant to FFM personality traits (Pennebaker and King 1999, Golbeck et al. 2011).

Furthermore, the acceptance rate of recommendations highly depends on the level of users' knowledge in a particular domain, we propose an approach to capture users' knowledge. For example, assume Mark has personal interests in cooking and music topics, and he is a music expert with a low level of knowledge in cooking. He is less likely to be influenced by other users' music suggestions or opinions, while he is more willing to seek other users' recommendations related to cooking topics as he has no expertise in this domain. Therefore, in addition to a user's personal inserts, his level of knowledge is another important factor, which is required to be considered in RSs. Integrating these three factors, users' personality type, their personal inserts and level of knowledge in a particular domain, into pure matrix factorization model can have a significant effect on the performance of recommender system. In this paper, we present novel algorithms and techniques to enable RSs discover the personality of a user and come up with suggestions with no need to analyze commonly rated items. The unique contributions of this paper are:

- To the best of our knowledge, this is the first personality-based recommender system considering users' personal interests and their level of knowledge in a particular domain, simultaneously.
- We develop an algorithm to construct a user-item interaction matrix with the actual ratings score to display the level of users' personal interests on items.
- We propose a novel Recommender System called APAR , integrating three main factors which may affect a user's decision making process, users' personal interests, users' personality type and their level of knowledge.
- We detect users' personality type implicitly with no need to users' efforts and through analyzing their online-generated contents.

- Experimental results on a real-world dataset demonstrate the effectiveness of our approach, especially in DSW-n-FCI situations.

The rest of the paper is organized as follows: firstly we present the related work, followed by an overview and the framework of the proposed approach. Next, we present the results of the evaluation of the proposed approach before concluding the paper with remarks on future research directions.

## Related Works

This section presents the related works and discusses the added value of our approach compared to the state-of-the-art techniques.

### *Recommender Systems*

The pure Collaborative Filtering (CF) systems, methods of making automatic predictions about the interests of a user by collecting preferences from many users, suffer from the data sparsity and cold-start problems. Many efforts have been made to address CF models' problems. Some studies resort to incorporating additional information including users' contents, like CTR (Wang et al. 2011), integrating social network information, such as SR (Ma et al. 2011), and exploiting review texts to find hidden information from them, like TopicMF (Bao et al. 2014). In order to mitigate the data sparsity problem, there is another attempt that captures semantic correlation between users and items, especially when there are not any in common rated items to make an accurate recommendation in tag-based recommender systems (Chen.C et al. 2016).

The main difference between the problem that they focused on and DSW-n-FCI is the fact that their problem can be a sub-category of DSW-n-FCI, which is a more general problem. Additionally, although most of the existing approaches rely on users' ratings to discover their interests, there are some attempts like CAPH which discovers users' features from their reviews and then based on the hotels' features, provides personalized recommendations (Ya-Han et al. 2016). DirectRelations is another study with a focus on extracting aspects from users' reviews and then generating a rule in a rule-based system to help customers when they are making a purchase decision (Chen, Y. et al. 2017).

### *Personality-Based Recommender Systems*

In contrast to traditional CF approaches, users' personality types, which can explain the wide variety of human behavior, have inspired some recommender systems. Personality is a domain-independent term along with individuals in a wide range of domains, like music (Mairesse et al. 2007), movies, books (Rentfrow et al. 2011). Moreover, personality can be predicted from nonverbal behavior (Bera et al. 2017, Ghafari et al. 2018.b), posts and activities like written review texts from online social media (Schwartz et al. 2013, Beheshti et al. 2013, Yakhchi et al. 2018, Ghafari et al. 2018a). Due to strong correlation between personality traits and users' preferences, RSs have been adopted to incorporate personality characteristics into their model to not only help users with a diverse set of items (McNee et al. 2006) but also to provide a better group recommendation (Kompan et al. 2014, Recio-Garcia et al., 2009) and improve the accuracy of RSs in Music, Movies, e-learning and web searches (Hu et al. 2010, Paiva et al. 2017, Tkalcic et al. 2015, Yakhchi et al. 2017). Hu and Pu (Hu and Pu 2011) demonstrated the correlations between users' personality types and their interests on Music genres ; then, they provided users with personality quizzes to explicitly detect their personality traits and they used the Pearson correlation coefficient to understand how much these users are similar. They achieved significant improvements compared to CF approaches that only take ratings into consideration for similarity detection. TWIN is an example of a recommender system that estimates users' similarity in terms of their personality traits with the Euclidian distance (Roshchina 2012, Roshchina et al. 2015).

### *Personality and User's Preferences*

Personality is described as "consistent behavior pattern and interpersonal processes originating within the individual" (Burger 2011) that can be acquired either explicitly by filling a questionnaire or implicitly through observing users' behavioral patterns. From the psychology point of view, personality is an important factor which is able to explain "patterns of thought, emotion, and behavior" (Funder 2011). Moreover, one of the key properties of personality is the fact that it is a stable behavioral pattern which humans tend to show regardless of their situations. Furthermore, there are several personality traits models which can explain human behaviors, and among them, Five Factor Model has drawn more attention both in psychology and computer science research. We adopt FFM as our personality detection model which is "the dominant paradigm in personality research, and one of the most influential models in all of the psychology" (Vinciarelli and Mohammadi 2014). As it is clear in Table.1 based on FFM, people's personality types can be categorized into five main traits which is briefly called OCEAN (McCrae 2009):

1. "Openness to Experience: creative, open-minded, curious, reflective, and not conventional";
2. "Agreeableness: cooperative, trusting, generous, helpful, nurturing, not aggressive or cold";
3. "Extroversion: assertive, amicable, outgoing, sociable, active, not reserved or shy";
4. "Conscientiousness: preserving, organized, and responsible";
5. "Neuroticism (Emotional Stability): relaxed, self-confident, not moody, easily upset, or easily stressed".

There are several questionnaire types based on FFM model, NEO-Personality-Inventory Revised (NEO-PI-R, 240 items) for instance, in which the participants' personality types are revealed after they answer several questions (Costa and McCrae 1995). Although personality detection with questionnaires might reveal a better understanding of a user's personality, it is a tedious and time-consuming task and thus users may be unwilling to attend to it. In contrast, in implicit personality detection models, user's digital footprints, and their behaviors and actions could be analyzed with no extra burden on users (Azaria et al. 2016, Beheshti et al. 2018).

**Added value.** Most of the aforementioned personality-aware methods not only detect users' personality types explicitly through filling a questionnaire, which is an infeasible task in many real-world applications, but also ignore the differences between users' level of knowledge in different domains. Most of the existing methods exploit users' personality types to simply use it as a similarity measure, in contrast to these approaches in this paper, in order to pay more attention to this important factor, we directly take it into our mathematical model and propose a novel matrix factorization model. The intuition of our method is users' decision-making can be affected by three major factors: users' personal interests, users' personality, and their level of knowledge in a particular domain. Moreover, to the best of our knowledge, there is not any study in the literature to assess the capability of personality–based RSs to deal with Data Sparsity With no Feedback on Common Items (DSW-n-FCI) problem.

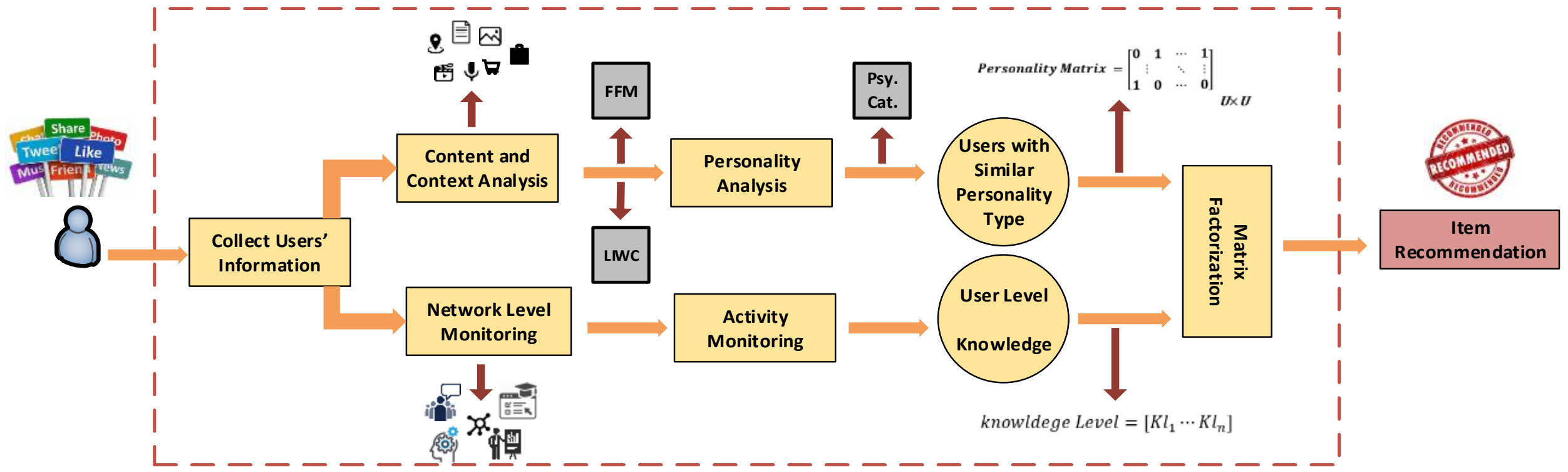

Figure 1. Our Analysis of Personality Aspects in Recommender Systems (APAR) Framework.

## Overview and Framework

Figure 1 depicts our novel framework called APAR. In this framework, "Psy. Cat." is extracted psychology categories from LIWC. For the first phase of this framework, we analyze the contents and contexts of users' generated information and contextualize these raw data to discover users' main characteristics, like their personality types. For the next phase, we monitor a user's activities and those of her neighbors' activities in a particular domain to figure out how other users are influenced by her opinions/comments to ascertain her level of knowledge. In the following sections, we give more details about each main phase of APAR.

Note that, our framework aims to capture users' interests in order to make recommendations more personalized, improve users' satisfaction, and boost business profits. To do that, we first construct a user-item interactions matrix with the real values of ratings in order to preserve the degree of users' interests and their personal interests. For example while it is true that between two users who gave 2 and 5 stars (in a Likert scale 0-5) to a specific item, they both may like this item, but one who gave 5 stars likes it much more. Then, we propose a novel matrix factorization model that incorporates users' personality type, users' level of knowledge and their personal interests.

## Problem Statement

Suppose there are *N* items $V = \{v_1, v_{2\ldots}\ v_N\}$, and *M* users $U = \{u_1, u_{2\ldots}\ u_M\}$. let $R \in R^{M \times N}$ represents the rating matrix, where $R_{ij}$ indicates ratings which have been given to item *i* by user *j* ($R_{ij} \in R^+$), and if the user has not seen this item, there are no available ratings for that item and we indicate this by assigning the *Null* value to $R_{ij}$. Let $L \in R^{M \times M}$ defines the personality matrix, where $L_{ij} = \{0,1\}$, and there is a direct correlation between $u_i$ and $u_j$ if they have the similar personality type. Most of the existing approaches consider that all the rated items are the same and if they have been rated by users, they will consider their ratings as 1. Thus, there is no difference between all the rated items while ratings values can indicate the actual level of users' personal interests on an item (He et al. 2017). In our approach, we construct user-item interactions matrix by using Eq.1 to show the level of users' interest on an item as follows (Xue et al. 2017):

$$w_{ij} = \begin{cases} 0, & if\ R_{ij} = Null \\ R_{ij}, & otherwise \end{cases} \quad (1)$$

, where $W \in R^{M \times N}$, and $w_{ij} > 0$ represents the interest of user *i* on item *j*, and $w_{ij} = 0$ represents no interest.

## Users' Personality Acquisition

Personality is a context-independent factor, which normally does not change over time and can be extracted explicitly, by using a questionnaire or implicitly by applying regression (Tkalcic et al. 2015). Unlike the most of existing studies, we detect users' personality types implicitly. People may also be different in their selected words because of their personality characteristics. Therefore, we collect all the written reviews of users to discover their personality types. Firstly, we use the Linguistic Inquiry and Word Count (LIWC) tool to understand how many words of users' reviews are related to each its 88 categories like *Positive emotions, Cognitive process,* and *Social processes*. Then, based on the psychology correlation between each category of LIWC and the Five Factor Model (FFM), we employ a linear regression model to measure a user's personality traits. In Eq.3, X, Y and Z represent LIWC categories and α, β, and γ are their weights in the relevant categories (Roshchina et al. 2012).

$$E = \alpha X + \beta Y + \gamma Z + \cdots \quad (2)$$

Then, we calculate *E* for each trait in FFM model. For example, let us consider a user with the conscientiousness personality type, it was shown in psychology that people with this personality type tend to use more words related to the Social Processes (SP) category, *e.g., talk, child,* Human Words (HW) category, *e.g, baby, man* and less Seeing Words (SW) category, *e.g., view, seen* (Cantador et al. 2013); then for this particular personality type, we have:

$$E = 0.264 * frequency\ of\ SP + 0.203 * frequency\ of\ HW - 0.227 * frequency\ of\ SW + \cdots \quad (3)$$

, where the weights are borrowed from (Cantador et al. 2013).

### *Users' Level of Knowledge*

The level of an individual's knowledge is one of the main criterions to determine the acceptance rate of his/her recommendations, which in this paper is termed it as users' level of knowledge. In the real-world, individuals may have different level of knowledge about various domains, but they may be an expert in one or some of them. We mark the level of knowledge of $u_i$ in domain $d$, as $kl_i^d$, and can be calculated as follows:

$$kl_i^d = {}^{1}/_{n_i^d} \sum_{p=1}^{n_i^d} h_p^{i,d} \qquad (4)$$

, where $n_i^d$ is the total number of reviews left by $u_i$ in that domain, and the ratings that are given to each review p by other users in this domain can be represented by $h_p^{i,d}$.

### *Our Model*

$u_i$'s interest on item $j$ in domain $d$ can be shown as an inner product between user $i$ latent feature vector $p_i^{(d)}$ and item $j$ latent feature vector $q_j^{(d)}$. Let matrix L contain personality information, and $\varphi_i^{+(d)}$ be the set of users who have the same personality type with user $i$. For instance, if $L_{ik} = 1$, it means $u_i$ and $u_k$ have the same personality type. For the sake of simplicity, we denote $\gamma_i^d = \beta + kl_i^d$ where parameter β controls the weight of users' preferences.

$$R_{ij}^{(d)} = \gamma_i^d p_i^{(d)^T} q_j^{(d)} + \left(1 - \gamma_i^d\right) \sum_{k \in \varphi_i^{+(d)}} L_{ik} P_k^{(d)^T} q_j^{(d)} \qquad (5)$$

In the above equation, $R_{ij}^{(d)}$ predicts the ratings values for unobserved items, which are based on the combination of three major parameters of users' personal interests, users' personality type, and their level of knowledge, into a unified model.

$$\min {}^{1}/_{2} \sum_{i=1}^{N} \sum_{j=1}^{M} I_{ij}^{(d)} \left( R_{ij}^{(d)} - \left( \gamma_i^d p_i^{(d)^T} q_j^{(d)} + \left( \left(1 - \gamma_i^d\right) \sum_{k \in \varphi_i^{+(d)}} L_{ik} P_k^{(d)^T} q_j^{(d)} \right) \right) \right)^2 + \alpha_1 \left\| P^{(d)} \right\|_F^2 + \alpha_2 \left\| Q^{(d)} \right\|_F^2 \qquad (6)$$

In the above equation, $\mathrm{I}_{\mathrm{ij}}^{(\mathrm{d})} = 1$, when user i has rated item j, otherwise $\mathrm{I}_{\mathrm{ij}}^{(\mathrm{d})} = 0$. In order to prevent overfitting, we introduce η(i, j) as the personality coefficient between $u_i$ and $u_j$ with some features 1) η(i, j) ∈ {0,1}, 2) η(i, j) = η(j, i) and 3) if η(i, j)=1, this means that $u_i$ and $u_j$ are more likely to have common interests. Then, similar to (Tang et al. 2013), we have $min \sum_{i=1}^{n} \sum_{j=1}^{m} \eta(\mathrm{i}, \mathrm{j}) \, \|P(i,:) - Q(j,:)\|_2^2$ as personality regularization. Hence, after some derivations for a specific user $u_i$, we have the following regularization:

$$\tfrac{1}{2}\sum_{i=1}^{n}\sum_{j=1}^{m}\eta(i,j)\|P(i,:)-Q(j,:)\|_2^2$$

$$=\tfrac{1}{2}\sum_{i=1}^{n}\sum_{j=1}^{m}\sum_{k=1}^{d}\eta(i,j)\left(P(i,:)-Q(j,:)\right)^2=\tfrac{1}{2}\sum_{i=1}^{n}\sum_{j=1}^{m}\sum_{k=1}^{d}\eta(i,j)P^2(i,k)+\tfrac{1}{2}$$

$$-\sum_{i=1}^{n}\sum_{j=1}^{m}\sum_{k=1}^{d}\eta(i,j)P(i,k)-Q(j,k)=\sum_{k=1}^{d}P^T(:,k)(D-Z)Q(:,k)=TrP^TYQ \quad (7)$$

,where Y=D-Z and Y is the Laplacian matrix, $D$ is a diagonal matrix with the $i_{th}$ diagonal element D (i; i) = $\sum_{j=1}^{n}Z(i,j)$, and Z is the personality coefficient matrix. Finally, our objective function can be represented as follows:

$$L^*=minP^{(d)}Q^{(d)}\,\tfrac{1}{2}\sum_{i=1}^{n}\sum_{j=1}^{m}I_{ij}^{(d)}\left(R_{ij}^{(d)}-\left(\gamma_i^d p_i^{(d)^T}q_j^{(d)}+\left(\left(1-\gamma_i^d\right)\sum_{k\in\varphi_i^{+(d)}}L_{ik}P_k^{(d)^T}q_j^{(d)}\right)\right)\right)^2+Tr(P^TLQ)+\alpha_1\left\|P^{(d)}\right\|_F^2+\alpha_2\left\|Q^{(d)}\right\|_F^2 \quad (8)$$

Next, with computing the partial derivations of L regarding to $P_i$ and $Q_j$, setting it to zero, and considering the KKT condition, we have:

$$P(i,j)\leftarrow P(i,j)\sqrt{\frac{A(i,j)}{B(i,j)}},\qquad Q(i,j)\leftarrow Q(i,j)\sqrt{\frac{C(i,j)}{D(i,j)}} \quad (9)$$

Note that, we omit complex mathematics here and provide the details of A, B, C and D in Appendix A.

Algorithm 1. The framework of APAR with Personality Regularization.

Input: L, d, γ, $\alpha_1$ and $\alpha_2$
Randomly initialize P, Q
**While** not converged **do**
    Set A as EQ.12 and Set B as Eq.13
    for j=1 to M do
        for i=1 to N do
            $P(i,j)\leftarrow P(i,j)\sqrt{\frac{A(i,j)}{B(i,j)}}$
        end for
    end for
    Set C as EQ.14 and Set D as Eq.15
    for j=1 to d do
        for i=1 to d do
            $Q(i,j)\leftarrow Q(i,j)\sqrt{\frac{C(i,j)}{D(i,j)}}$
        end for
    end for
    end for
**end while**
return P,Q

**End Algorithm**

## Experimental Settings and Analysis

In this section, we discuss the results of comprehensive experiments over a real-world dataset to compare the performance of our model with other state-of-the-art methods, we also examine APAR's efficiency in the presence of DSW-n-FCI problem.

**Dataset.** We evaluate our proposed model on the Amazon dataset, which has been widely used in recommender systems (Zheng et al. 2017). This dataset consists of 2000 of users, 1500 items, 86690 reviews, 7219 number ratings, 3.6113 average number of rates per user, 0.2166 average number of rates per item and user ratings density is 0.0024. In this paper, we employ a subset of Amazon dataset, Instant-videos, due to the strong correlation between users' preferences on video and their personality types. We will investigate the other domains in our future works. The dataset includes users who wrote more than 3 reviews; additionally, after applying a five-fold cross-validation, we set parameters as $\gamma = 0.5$, $d$=100, $\alpha_1 = 0.1$, and $\alpha_2 = 0.1$ in order to have the best recommendation accuracy.

**Evaluation.** In our experiments, we select two well-known evaluation metrics: Mean Absolute Error (MAE) and Root Mean Squared Error (RMSE). The smaller MAE and RMSE demonstrate a better recommendation accuracy;

$$MAE = \frac{\sum_{(u,i)\in R_{test}} |\hat{R}_{ui} - R_{ui}|}{|R_{test}|}, \qquad RMSE = \sqrt{\frac{\sum_{(u,i)\in R_{test}} |\hat{R}_{ui} - R_{ui}|}{|R_{test}|}} \qquad (10)$$

,here $R_{ui}$ and $\hat{R}_{ui}$ are the real and estimated ratings values, respectively, and $R_{test}$ represents total number in the test dataset.To evaluate the performance of our proposed approach (APAR), we compare it with the following three categories of recommenders: (i) purely rating based recommenders; in this category we select SVD++, utilizing user-item interactions for recommendations (Koren et al. 2008), UserMean and ItemMean, measuring the mean of all ratings that a user has given to items and the mean of all ratings that an item has been given by users, respectively; and Random method that randomly assigns rated values. (ii) Integrating users' contents; we compare our model with the-state-of-the-art model in this category like CTR, using topic molding to combine the contents of documents with traditional collaborative filtering in order to recommend scientific articles. (iii) Personality-based recommender systems; most of the previous studies in this category detect users' personality types explicitly.

We select TWIN and Hu for our comparisons, as TWIN recognizes users' personality types explicitly and then recommends hotels to users with similar personality types and Hu as the explicit personality recognition approach defines a new similarity detection metric in music domain that is a linear combination of both personality and rating information. Although both TWIN and Hu and Pu models detect users' personality types explicitly, to implement them in this paper, we used their own extracted weight of correlation between categories of LIWC tool and FFM personality model. In contrast to above personality-based approaches, our model is a general purpose recommender system which not only detect users' personality types implicitly but also take users' personal interests and their level of knowledge into account.

### *Performance Comparison and Analysis*

In this section, we compare the performance of our proposed method with other approaches regarding RMSE and MAE. As it can be seen in Table.1, we use different sets of the training data size (60%; 70%; 80%; 90%). Table.1, demonstrates that while we increase the volume of the training data, the accuracy of all models will be improved, accordingly. All the approaches, reached their best performance in both metrics when they were trained with 90% of the data. Therefore, to have a fair comparison, we consider the results related to 90% training size. Among all compared methods, our proposed model, APAR, showed the best performance in terms of both RMSE and MAE. As it can be observed from Table.1, SVD++ performs better than UserMean and ItemMean, because they simply use the mean rating value of a user and an item, respectively.

| Training Data | Metrics | Random | Hu | ItemMean | UserMean | SVD++ | CTR | TWIN | APAR |
|---|---|---|---|---|---|---|---|---|---|
| 60% | MAE | 7.759 | 2.983 | 2.889 | 2.770 | 2.033 | 1.887 | 1.353 | 1.165 |
| 60% | RMSE | 8.124 | 3.32 | 3.084 | 2.896 | 2.314 | 2.129 | 2.074 | 1.554 |
| 70% | MAE | 7.467 | 2.763 | 2.648 | 2.54 | 1.864 | 1.69 | 1.215 | 1.005 |
| 70% | RMSE | 7.985 | 3.152 | 2.815 | 2.785 | 2.29 | 2.101 | 1.898 | 1.295 |
| 80% | MAE | 7.058 | 2.581 | 2.354 | 2.321 | 1.719 | 1.52 | 1.132 | 0.936 |
| 80% | RMSE | 7.654 | 2.978 | 2.642 | 2.763 | 2.011 | 1.986 | 1.620 | 1.058 |
| 90% | MAE | 6.589 | 2.426 | 2.224 | 2.203 | 1.547 | 1.391 | 0. 95 | 0.850 |
| 90% | RMSE | 7.236 | 2.649 | 2.545 | 2.539 | 1.99 | 1.718 | 1.428 | 0.995 |

**Table 1: Performance analysis on the Amazon dataset.**

However, the recommendation performance of UserMean is higher than that of ItemMean, because the average number of ratings per user is more than the average number of ratings per item, in this dataset. Moreover, the recommendation performance of CTR is higher than that of SVD++ by around 11% and 14% in terms of MAE and RMSE respectively, since SVD++ only takes user's ratings into consideration. In the personality-based recommender systems, TWIN performs better than Hu, because Hu integrates both ratings and personality information and in because of possible lack of available ratings in real-world scenarios, it may face with a performance degradation. Finally, APAR which detects users' personality types implicitly, has a superior performance compared to other methods because it integrates users' personality types, their personal interests, as well as their level of knowledge. In addition, unlike the majority of existing approaches, APAR constructs a matrix with actual ratings score which can help to better understanding users' interests. It outperforms the recommendation performance of CTR by 39%, 42% and TWIN by 11%, 31% in terms of MAE and RMSE respectively.

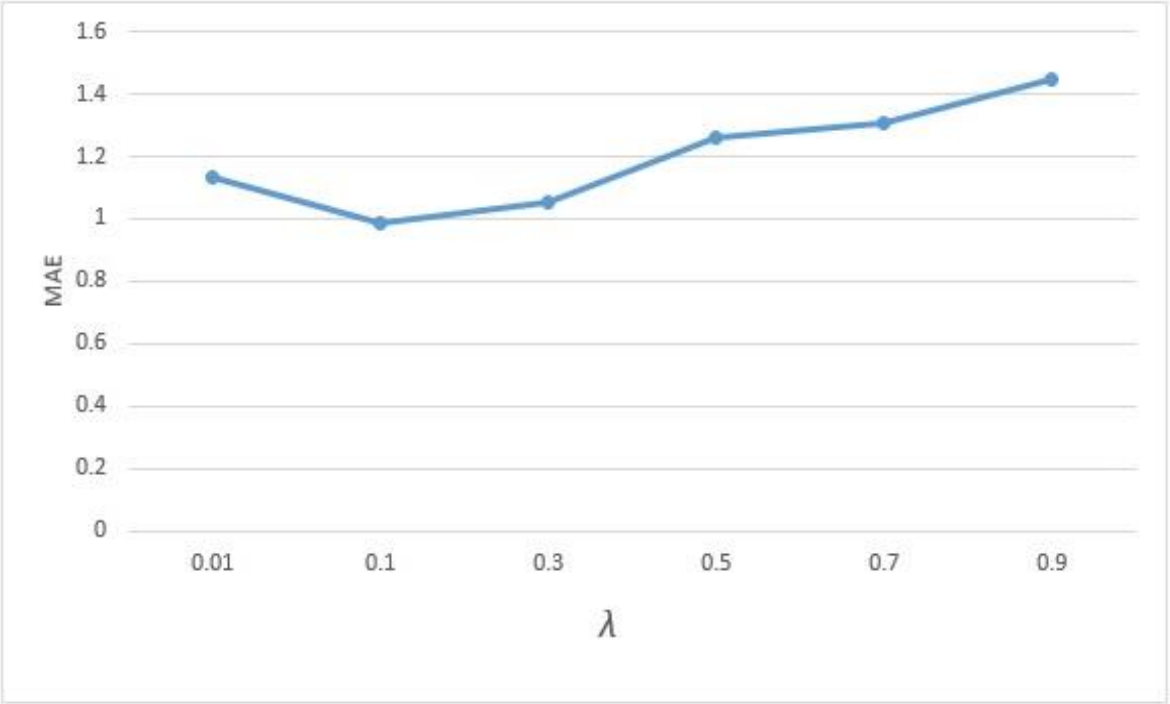

**Figure 2(a). Regarding to MAE.**

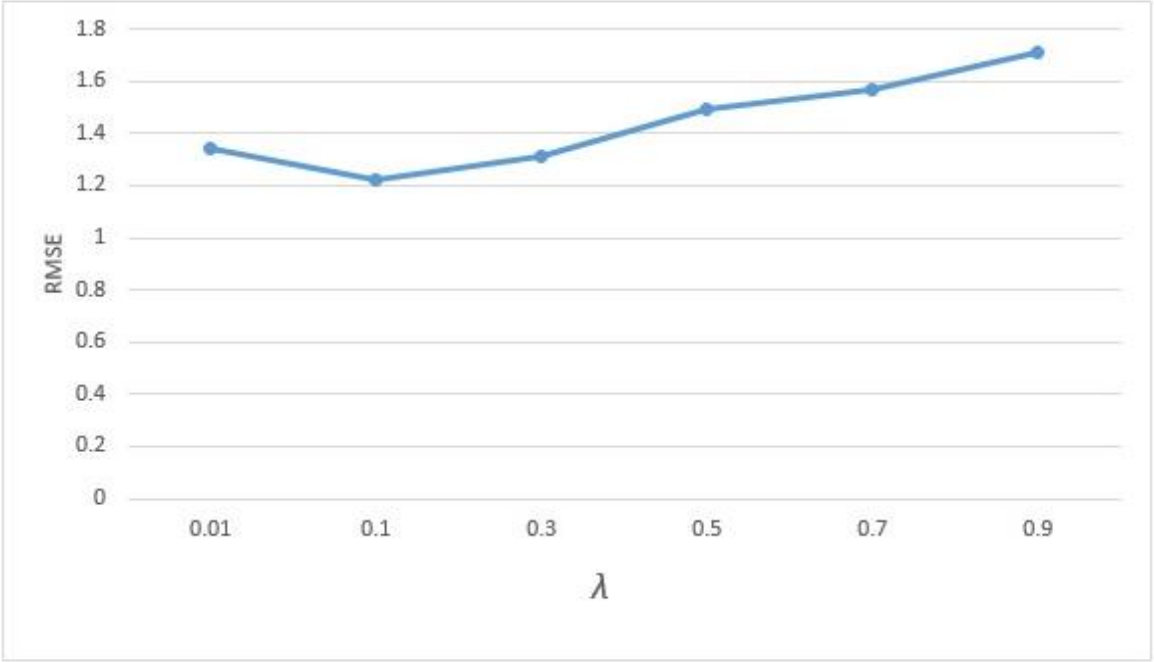

**Figure 2(b). Regarding to RMSE.**

**Figure 2. Impact of Personality Regularization.**

**Analysis and Summary.** The experimental results demonstrate that our proposed model performs better in both MAE and RMSE metrics compared to the other two models. APAR pays more attention to the real values of ratings, which can explain the level of interests of a user for an item. In addition, it considers users' personality types and their level of knowledge in addition to their personal interests.

### *Impact of Personality Regularization*

In this section, we explore the impact of our personality regularization factor on the effectiveness of recommendations regarding RMSE and MAE, by considering λ as a controlling parameter. Therefore, we assigned different values to λ as {0.01, 0.1, 0.3, 0.5, 0.7, and 0.9} and the results are shwon in Fig.3. The best performance of APAR is achieved, when λ increased. In this case, MAE and RMSE start to decrease and recommendation performance increases, subsequently. At that point, with the growing value of λ, the values of MAE and RMSE also increase and accordingly we observed performance degradations. We compute the average of all the MAE and RMSE values at different training data sizes. As it can be seen in Fig.2, our approach achieved its best performance (MAE = 1 and RMSE = 1.2) when λ = 0.1.

**DSW-n-FCI Experiments.** We do the following experiments to examine the recommendation performance of these methods while dealing with DSW-n-FCI problem. In order to show APAR's capability to overcome the Data Sparsity With no Feedback on Common Items (DSW-n-FCI) problem, we randomly divide the dataset into four different sub-datasets with the DSW-n-FCI degree of 20% (SD1), 40% (SD2), 60% (SD3), and 80% (SD4). When the degree of DSW-n-FCI is 20%, it means only 20% of users do not have any feedback on common items out of all users. We compute the degree of DSW-n-FCI as follows:

$$d\% = \frac{number\ of\ users\ with\ no\ feedback\ on\ common\ items}{total\ users} \qquad (11)$$

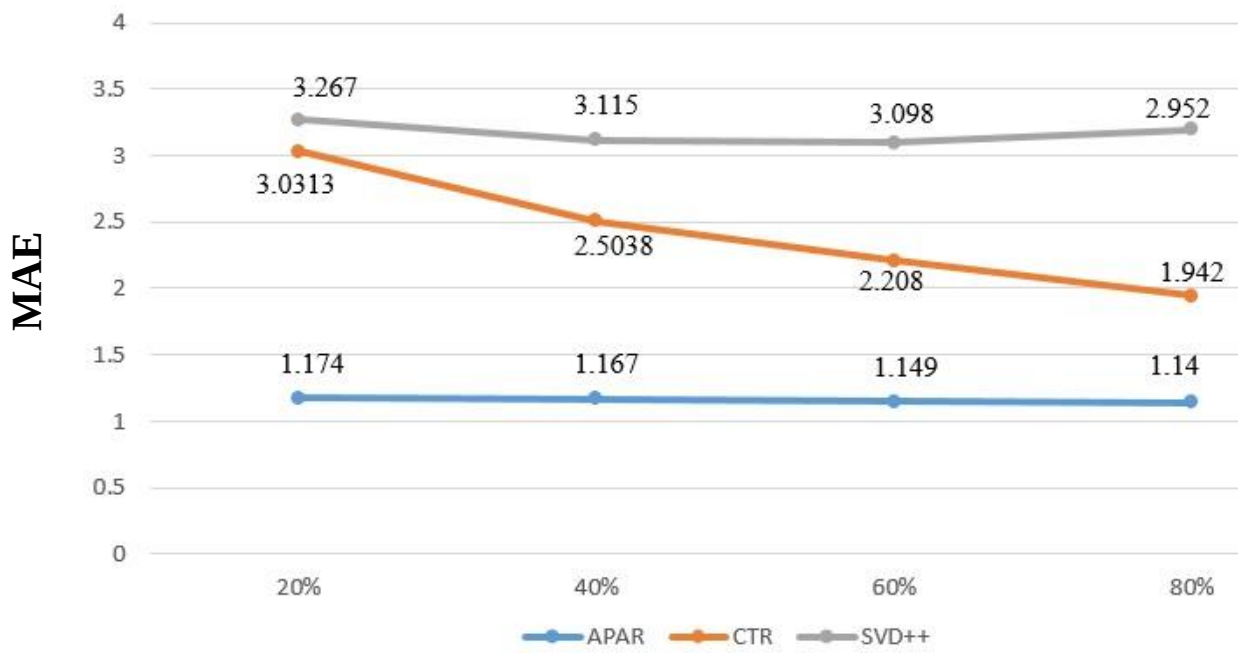

**Figure 3 (a). Regarding to MAE.**

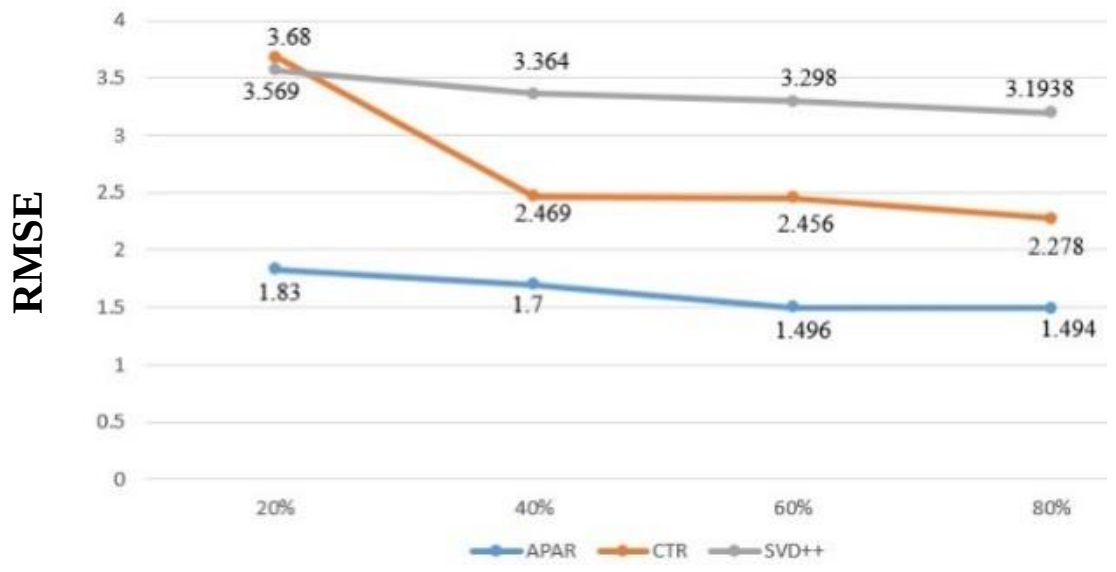

**Figure 3 (b). Regarding to RMSE.**

**Figure 3. Recommendation Performance in Presence of DSW-n-FCI.**

Then, we compare our model with other methods on each sub-dataset explained in the previous section. As it can be seen in Fig.3, APAR, achieves the best performance in all the sub-datasets and different degrees of DSW-n-FCI. We compare APAR with CTR, and SVD++ since each of the methods shows great performance in our experiments. In SD1, SD2, SD3 and SD4 sub-datasets, APAR was successful to boost performance on average by 51%, 63% and 39%, 52% compared to CTR and SVD++ in terms of MAE and RMSE, respectively.

**Summary**. The experimental results show the superior performance of APAR to handle the DSW-n-FCI problem. It is observed from Figure 3 that the efficiency of APAR which used personality information is able to make a recommendation even when there is no any feedback on in common items.

## Conclusions and Future Work

In this paper, we first have introduced the DSW-n-FCI problem which recommender systems are confronted in real-world scenarios. Then, with this assumption that the real value of ratings can be a good indicator of the level of users' personal interests and we a construct a user-item interaction matrix accordingly. Next, in order to solve the DSW-n-FCI problem, we enriched our model by incorporating the users' personality types without any need for users' extra efforts through their written reviews. Since different users with a similar personality type may have a different level of expertise in a particular domain, we also took the users' level of knowledge into account as another key factor. Finally, with these three factors which can influence our decision making process, we proposed a novel matrix factorization model. The experimental results on a real-world dataset demonstrate that our model outperforms some of the state-of-the-art models regarding RMSE and MAE, especially in DSW-n-FCI situations. For our future work, we will focus on a cross-domain personality-based recommender system to enrich our available information and mitigate the data sparsity problem.

## Appendix A

Below, we represent A, B, C, and D which are part of the updating rules in Equation 9 that defied in our model. We omit them in paper and represent them here due to simplification and avoid more complexity:

$$A = q^{(d)^T} R^{(d)} \gamma^{(d)} + \left(1 - \gamma_i^{(d)}\right) L^T R^{(d)^T} q^{(d)^T} + \gamma^{(d)} R^{(d)} q^{(d)^T} \quad (12)$$

$$B = \gamma^{(d)} q^{(d)} p^{(d)} q^{(d)^T} + q^{(d)^T} \gamma^{(d)} p^{(d)} q^{(d)} + \gamma^{(d)} p^{(d)} L^T q^{(d)^T} + \left(1 - \gamma_i^{(d)}\right) L^T q^{(d)^T} p^{(d)^T} + \left(1 + \gamma_i^{(d)}\right) L^T R^{(d)^T} q^{(d)^T} + \left(1 - \gamma_i^{(d)}\right) L^T q^{(d)} p^{(d)} q^{(d)^T} + \left(1 - \gamma_i^{(d)}\right) q^{(d)^T} L p^{(d)} q^{(d)} + \left(1 - \gamma_i^{(d)}\right) L^T q^{(d)^T} p^{(d)^T} + YP + Y^T p + \alpha_1 2p \quad (13)$$

$$C = R^{(d)} \gamma^{(d)} p^{(d)^T} + \left(1 - \gamma_i^{(d)}\right) p^{(d)^T} L^T R^{(d)^T} + \gamma^{(d)} p^{(d)^T} R^{(d)} \quad (14)$$

$$D = \gamma^{(d)} p^{(d)^T} q^{(d)} p^{(d)} + \gamma^{(d)} p^{(d)} q^{(d)} p^{(d)^T} + \left(1 - \gamma_i^{(d)}\right) p^{(d)^T} q^{(d)^T} L^T + \left(1 - \gamma_i^{(d)}\right) p^{(d)^T} L^T q^{(d)^T} + \left(1 + \gamma_i^{(d)}\right) p^{(d)^T} L^T R^{(d)} + \left(1 - \gamma_i^{(d)}\right) p^{(d)^T} L^{(T)} q^{(d)} p^{(d)} + \left(1 - \gamma_i^{(d)}\right) L p^{(d)} q^{(d)} p^{(d)^T} + \left(1 - \gamma_i^{(d)}\right) p^{(d)^T} L^{(T)} q^{(d)^T} p^{(d)^T} L^{(T)} + \left(1 - \gamma_i^{(d)}\right) p^{(d)^T} L^{(T)} q^{(d)^T} p^{(d)^T} L^{(T)} + \alpha_2 2q \quad (15)$$